# An integrated view of Quantum Technology? Mapping Media, Business, and Policy Narratives

Viktor Suter
Institute for Media and
Communications Management
University of St. Gallen
viktor.suter@unisg.ch

Charles Ma
Institute for Media and
Communications Management
University of St. Gallen
charles.ma@unisg.ch

Gina Pöhlmann
Institute for Media and
Communications Management
University of St. Gallen
gina-maria.poehlmann@unisg.ch

Miriam Meckel
Institute for Media and
Communications Management
University of St. Gallen
miriam.meckel@unisg.ch

Lea Steinacker
Institute for Media and
Communications Management
University of St. Gallen
lea.steinacker@unisg.ch

**Abstract**

*Narratives play a vital role in shaping public perceptions and policy on emerging technologies like quantum technology (QT). However, little is known about the construction and variation of QT narratives across societal domains. This study examines how QT is presented in business, media, and government texts using thematic narrative analysis. Our research design utilizes an extensive dataset of 36 government documents, 165 business reports, and 2,331 media articles published over 20 years. We employ a computational social science approach, combining BERTopic modeling with qualitative assessment to extract themes and narratives. The findings show that public discourse on QT reflects prevailing social and political agendas, focusing on technical and commercial potential, global conflicts, national strategies, and social issues. Media articles provide the most balanced coverage, while business and government discourses tend to overlook societal implications. We discuss the ramifications for integrating QT into society and the need for well-informed public discourse.*

## 1. Introduction

Quantum computing represents a significant shift from classical computing. Classical computers use bits that are either 0 or 1, while quantum computers use qubits, which can be 0, 1, or both simultaneously due to superposition (Rietsche et al., 2022). Additionally, entanglement allows qubits to be correlated, providing information about each other even over long distances (Ding & Chong, 2020). These principles enable quantum technology (QT) to potentially revolutionize data processing, secure communications, and computational capabilities. Although QT is still experimental, practical approaches combining classical and quantum systems are emerging in defense, finance, and healthcare (Flöther, 2023; Egger et al., 2020; Krelina, 2021). Furthermore, researchers in chemistry and materials science are applying quantum computers to optimization problems or complex simulations of chemical processes, often integrating it with machine learning and big data analytics (Schuhmacher et al., 2022). Governments and companies see QT as a strategic asset, with significant funding from both public and private sources, including Google, Huawei, IBM, and venture-backed startups (Gibney, 2019). This transition of QT to real-world applications is fraught with high expectations. Although quantum computers are currently used for a limited number of specific tasks, mainly for high-precision simulations of, e.g., molecules and small particles (Daley et al., 2022), they promise to help solve complex scientific problems. Some see this technology as a solution to climate change and public health issues, enabling innovative battery designs (Downing & Ukhtary, 2023) or accelerating drug discovery (Flöther, 2023). The key aspect here is that understanding molecular structures is critical to many industries, and quantum computers are expected to solve problems that current classical systems cannot. Others, however, view the technology as a threat. Experts predict that a commercially viable quantum computer - which is still years away - might have the potential to break existing encryption schemes, posing risks to privacy, cybersecurity and societal trust (Gasser et al., 2024).

The public debate about QT mirrors past discussions about artificial intelligence (AI), highlighting issues like job displacement, economic impacts (Brynjolfsson & McAfee, 2016), privacy, surveillance (Zuboff, 2019), and ethical concerns (O'Neil, 2016). Similarly, QT is seen as both a promise for innovation and a potential risk.

We are interested in the competing public narratives about QT constructed by the media, governments, and business stakeholders, as these narratives significantly influence the perception and adoption of technology (Gerhards & Schäfer, 2009; Malone et al., 2017). For instance, AI narratives have set expectations that have impacted regulatory responses and public trust (Cave et al., 2018; Cave & Dihal, 2019). To explore this, we collected textual data from the past 23 years, including 36 government policy documents, 165 business reports, and 2,331 news articles. We used a mixed-methods approach, combining BERTopic modeling with qualitative coding and validation, to extract topics, group them into higher-level themes, and identify prevalent narratives. This approach is situated in the literature on AI and QT narratives, providing a conceptual foundation for our analysis. Our research aims to answer the following questions:

(1) What narratives about quantum technology do policymakers, corporations, and the media construct?
(2) How do these narratives compare and contrast across these sources?
(3) How has QT discourse changed over time?

This study is structured as follows: The first section introduces the concept of narratives and reviews previous findings using AI as an example, explaining their relevance to QT discourse. The next section outlines our analytical approach, including data collection and processing procedures. Finally, we present the findings, followed by a discussion and concluding remarks.[1]

## 2. Narratives about AI and QT

In the social sciences, foundational literature (Bruner, 1991; Polkinghorne, 1995) offers two main definitions of narrative. The first broadly views narrative as structured communication in natural language, essentially textual data. The second characterizes narrative as discourse that organizes human experience into meaningful patterns, enabling individuals to connect disparate facts into a coherent story. This view highlights narratives' role in providing individuals and collectives with a shared interpretation of reality (Patterson & Monroe, 1998) and their political dimension when shared understandings are challenged. Research in science and technology studies has focused on how narratives shape future expectations and frame technological innovation. This includes how visionary rhetoric in corporate settings influences economic dynamics by generating expectations that guide business strategies and investment decisions (Beckert, 2016), how "socio-technical imaginaries" affect governmental agendas and public fund allocation (Jasanoff & Kim, 2009, 2015), and how this is reflected in media coverage. This research demonstrates the role of narratives about technologies in shaping public perceptions and opinions. Our analysis is grounded in this conceptualization of narratives.

### 2.1. Learning from Narratives about AI

Narratives about AI serve as a blueprint for the discussion and concerns that are beginning to surface around QT. As QT moves from theoretical research to practical applications, it may follow a similar trajectory to AI, with narratives ranging from utopian transformative potential to fears of expanding surveillance capabilities or an accelerating global arms race (Nouwens & Legarda, 2018). Moreover, debates about the societal implications of QT are beginning to reflect concerns about equity and accessibility similar to those discussed in the context of AI (Lutz, 2019).

Past studies provide an overview of fictional and non-fictional portrayals of AI, arguing that popular narratives tend to be framed around exaggerated hopes or melodramatic fears (Cave et al., 2018; Cave & Dihal, 2019). They also draw parallels with earlier emerging technologies and argue for the necessity of diverse and realistic narratives about AI to better inform public debate and policy discussions.

Narratives developed in policy contexts are particularly relevant to understanding how texts shape real-world futures. National AI strategy documents actively contribute to creating the conditions they describe, effectively "talking AI into being" (Bareis & Katzenbach, 2022). Critical voices argue that this dynamic often serves powerful industry interests at the expense of broader societal good (Veale et al., 2023).

Analyses of media coverage show that AI narratives often overlook its limitations, creating a gap between public perceptions and actual technological capabilities (Chubb et al., 2022; Elish & Boyd, 2018). Dramatized depictions of AI in science fiction can also mislead the public, potentially distorting perceptions and influencing regulatory responses (Hermann, 2023). Powerful systems like ChatGPT encourage people to attribute human-like intelligence to AI, feeding myths of impending AI omnipotence (Walsh, 2023) and distracting from present-day challenges posed by AI (Crawford, 2021).

---
[1] Parts of this research were presented in a workshop at the IEEE Quantum Week 2024 conference.

## 2.2. The New Narratives about QT

The literature on QT narratives is less extensive than that on AI, with only a few studies focusing directly on narratives. Grinbaum (2017) notes that communication about QT has been divided into pragmatic and theoretical narratives. Meinsma et al. (2023) show that TEDx talks on QT focus on the "spooky" aspects of quantum physics and tend to emphasize the technology's benefits over risks. Godoy-Descazeaux et al. (2023) and Hilkamo and Granqvist (2022) study metaphors used to describe QT, highlighting the novelty and "weirdness" of the technology.

Discussions about democratization focus on making QT accessible and enabling citizens to participate in decisions about innovation progress (Seskir et al., 2023). However, early access to QT is limited to wealthy Western nations, large global technology companies, and China, potentially leading to considerable global geographic disparities (Ten Holter et al., 2022). Because the potential of QT is still largely theoretical, there has been limited ethical debate about it. One analysis suggests that for QT applications to be beneficial beyond actors involved in industrial or geopolitical competition, they must be culturally embedded with outreach efforts that reflect diverse values of different stakeholders (Coenen et al., 2022). Concerns have also been raised about the narrow framing of societal progress in QT research (Roberson, 2021).

The governance debate on QT in academic literature tends to focus on responsible innovation frameworks that promote ethical standards, sustainability, and socially desirable outcomes (Inglesant et al., 2021; Coenen & Grunwald, 2017). Other proposed frameworks integrate ethical, legal, socio-economic, and political dimensions into QT research and development (Gasser et al., 2024). In addition, legal experts advocate for flexible regulation that balances innovation and risk mitigation while considering national security imperatives (Johnson, 2019).

## 3. Methods and Data

Methodologically, our work employs thematic narrative analysis, which focuses on identifying themes and narratives within texts (Braun & Clarke, 2006; Riessman, 2008). This approach, a variant of content analysis, captures surface-level semantic content through theme identification and explores narratives related to respective themes for broader interpretive commentary.

For this analysis, we use BERTopic modeling to extract topics, apply qualitative coding to group these into higher-level themes, and then highlight the narratives contained in the data. The specific steps involved are outlined below.

### 3.1. Data Collection

We collected data from three different sources: business, media, and politics. For the business component, we obtained publicly available reports from the world's 15 largest consulting firms by revenue (McCain, 2023) and consulted the Swisscovery database for relevant articles published between January 1, 2002 and January 31, 2023.

For the media component, we collected English-language articles from eight national newspapers in the US, UK, China, and India, as these countries play a key role in shaping the QT landscape (Kung, 2022). The newspapers included were: The Wall Street Journal, The Washington Post (United States), The Daily Telegraph, The Guardian (United Kingdom), The Times of India, The Economic Times (India), and China Daily, South China Morning Post (China). The data collection period covered articles published between January 1, 2000 and August 15, 2023.

To collect our sample of policy documents, we searched government websites that mention policy publications, such as Canada (Innovation, Science and Economic Development Canada, 2023) Germany (Bundesministerium für Bildung und Forschung, 2018), India (Ministry of Science and Technology, 2024), and the United States (Office of Science and Technology Policy, 2024). We also consulted technology blogs, such as The Quantum Insider[2], to identify new policy publications. It is worth noting that the data for policy and business sectors is predominantly concentrated after 2015/2016, limiting our analysis of these sectors to more recent years. This difference in data volume and time span should be considered when interpreting longitudinal trends.

### 3.2. Data Cleaning and Preprocessing

We collected business reports and policy documents as PDF files, which we cleaned by removing headers, footers, acknowledgements, author information, citation notes, glossaries, bibliographies, and appendices. After converting them to txt files, we used RStudio for further text cleaning, resulting in 165 business documents and 36 government policy documents. For the media category, we initially retrieved 2,754 articles. We ensured relevance to QT through keyword searches, excluded articles unrelated to QT or longer than 10,000 words, and removed duplicates and highly similar articles using a Levenshtein distance cutoff value of 0.3 (Van der Loo, 2014). After further cleaning, we obtained a final sample of 2,331 media articles.

---
[2] https://thequantuminsider.com

### 3.3. Topic Modeling

We applied BERTopic, a new topic modeling method based on deep learning transformer models, to explore the underlying topics and narratives in the documents (Grootendorst, 2022a). BERTopic uses dense clustering algorithms and class-based term frequency-inverse document frequency (c-TF-IDF) scores to generate coherent topics, resulting in higher quality and more interpretable topics compared to traditional topic modeling techniques such as LDA (Blei et al., 2003) and STM (Roberts et al., 2019). We split each document into individual sentences and fed them into BERTopic, resulting in 22,005 sentences from 165 business-focused documents, 67,012 sentences from 2,331 media articles, and 14,806 sentences from 36 government policy documents. We used the all-distilroberta-v1 sentence embedding model (Reimers, 2022), UMAP for dimensionality reduction, and HDBSCAN for clustering. We tuned the HDBSCAN parameters using TopicTuner (Drob-Xx, 2023) and manual inspection (Grootendorst, 2021). After parameterization and post-processing, we obtained 51 topics in the business category, 55 topics in the media category, and 51 topics in the government policy category.

### 3.4. Determining Themes & Narratives

The resulting topics are diverse, but often focus on similar concepts. Grouping topics into higher-level themes and identifying narratives provides a holistic analysis that reveals cross-cutting relationships. Themes are patterns that organize and summarize data points, while narratives illustrate specific stories within these themes (Riessman, 2008). We followed Braun and Clarke's (2006) and Riessman's (2008) qualitative research process for identifying themes and narratives.

We initially identified six themes, created a codebook, and conducted multiple rounds of coding with three independent coders. We calculated Krippendorff's Alpha values for intercoder agreement and refined the codebook to five themes. Krippendorff's Alpha values for each of the five themes reached at least 0.8, meeting the standard for good inter-coder reliability (Krippendorff, 2019).

Finally, we identified narratives by revisiting each theme's topics and representative sentences from our BERTopic results. To support the credibility of this interpretative step, we provide descriptions and text excerpts for each theme and its associated narratives.

## 4. Results

From the process described in the previous section, we identified five overarching themes: Technical Aspects & Applications, Politics & Global Conflicts, People & Society, National Technology Strategies, and Business & Market Development. These themes represent the key areas of discourse surrounding QT. To answer our research questions, we describe each theme and associated narratives in turn.

### 4.1. What narratives about quantum technologies do policymakers, corporations, and the media construct?

The "Technical Aspects & Applications" theme covers the fundamental technical details, mechanics, and practical implications of emerging QT across applications domains and industries, including qubits, cryptography, AI, drug development, and logistics. Texts within this theme often portray QT as overcoming the limitations of classical computing by exploiting quantum mechanical principles. The theme also captures how discussions about technical foundations are closely connected to and mentioned alongside practical uses of QT, as the following narratives will show.

One characteristic narrative associated with this theme is that of "Revolutionary Technological Advancement". Here, QT is described as a radical force that will transform the technological landscape and enable substantial progress across businesses and industries. The following excerpt is exemplary of this narrative:

> *"Quantum technology, which enables the manipulation of atoms and sub-atomic particles, will allow for a new class of ultra-sensitive devices with key potential to profoundly impact and disrupt significant applications in areas such as defense, aerospace, industrial, commercial, infrastructure, transportation and logistics markets."* (From Business & Industry, Frost & Sullivan, 28.03.2020)

Another popular narrative is "Challenges to Traditional Cybersecurity". This narrative references, among other things, Shor's algorithm, a powerful quantum algorithm that can potentially break any existing classical cryptographic method used to secure digital communications. It posits that as QT is increasingly integrated into practical applications, traditional data security, privacy, and overall cybersecurity frameworks are at risk. Consequently, this narrative includes calls for the development of "post-quantum" cryptographic techniques that are immune to quantum-based intrusions. The following representative sentences illustrate this point:

> *"Quantum computers have the potential to break existing encryption security algorithms; governments and organizations alike must therefore ensure encryption is quantum-ready."* (From Business & Industry, Accenture, 05.03.2021)

> *"This motivates the development of post-quantum cryptography, i.e. encryption methods that quantum computers could not break."* (From Government Policies, Quantum Manifesto, EU, 2016)

A third common narrative is "Quantum Mysteries and Magical Phenomena". This narrative addresses the puzzling aspects of QT that defy everyday intuitions about the behavior of objects around us. It typically refers to concepts like entanglement, famously described by Einstein as "spooky action at a distance," the creation of high-fidelity magic states in quantum computing, and the paradoxical nature of Schrödinger's cat thought experiment. Invitations to rethink the nature of reality itself, the limits of computation, or the future of communication technology are often included in this narrative. Consider these representative excerpts:

> *"Circuits to create high-fidelity magic states are known as magic-state factories, and the process of creating these states is known as magic-state distillation."* (From Business & Industry, Blunt et al., 2022)

> *"Entanglement – once described by Einstein as 'spooky action at a distance' – is the phenomenon that will underpin a (future) quantum internet's power and fundamental security."* (From Government Policies, National Agenda for Quantum Technology, Netherlands, 2019)

The "Politics & Global Conflicts" theme encompasses geopolitical dynamics, and international relations. It addresses the interactions and tensions between different nation-states and world regions and examines how QT intersects with the challenges and opportunities that arise in these relationships. The theme discusses QT not only as an advancement in computational capabilities, but also, and chiefly, as a central element in power relations between nations.

A widely circulating narrative is the "Sino-American Rivalry", which reflects the competition between the United States and China for technological supremacy. It includes accounts of the Biden administration's actions to impose sanctions and investment restrictions on China's QT and AI sectors to limit China's technological advances and protect U.S. national security. With Beijing vowing to retaliate, the narrative also often addresses the decoupling of technology research and development between these major political and economic powers. The following excerpt illustrates the rhetoric and actions associated with this narrative:

> *"President Joe Biden signed an executive order on Wednesday to block US dollars from flowing to Chinese semiconductors and microelectronics, quantum information tech and certain artificial intelligence systems- the latest move to blunt China's access to such technologies."* (From News Coverage, South China Morning Post, 08.11.2023)

A more general but related narrative is "QT in Global Power Dynamics". It addresses the role of QT in international conflicts and alliances beyond the Sino-American rivalry. QT is depicted as a bargaining chip in geopolitical strategies such as imposing economic sanctions to curb technological progress, enforcing technology embargoes to restrict access, and promoting cooperative efforts to share QT innovations with allies. The narrative recounts how QT is deployed as part of an international relations and diplomacy playbook that focuses not only on competition and innovation for prosperity, but on gaining high-tech advantage and political influence.

> *"The growing realisation of the emerging opportunities has already kick-started an international race to turn the recent excellent achievements of quantum science into a national competitive advantage in Quantum Technologies."* (From Government Policies, Positioning Ireland for the quantum opportunity, Ireland, 2019)

The "People & Society" theme captures the potential impact of QT on societal issues, long-term sustainable development, and how the technology can benefit society as a whole. It includes issues such as gender inequality or responsible technology development.

One of the main narratives tied to this theme is "Education Reform and Workforce Development". It presents QT as requiring that education and training needs in QT fields be addressed and emphasizes the need for learning environments capable of producing a quantum-ready workforce. Calls for integrating QT into curricula at all levels to foster a skilled and receptive workforce are part of this narrative. Consider this excerpt from an EU policy document:

> *"The creation of a learning ecosystem embracing the concepts of quantum physics at all levels ranging from school up to the working environment is required, not just for a quantum-ready workforce to emerge, but for a well-informed society with knowledge and attitudes towards the acceptance of quantum technologies."* (From Government Policies, European Quantum Flagship, EU, 2020)

Another popular narrative within this theme is "Promoting Gender Equality", which draws attention to issues of gender balance in QT research and development activities. The narrative points out that female researchers are often at the margins of research and technology development. Some documents also explicitly advocate for the adoption of norms and policies to address disparities in gender visibility and representation, such as the European Quantum Flagship initiative below. The following excerpts provide an example:

*"Gender equity in conferences: through the implementation of the conference charter and the promotion of the visibility of women in quantum, Quantum Flagship related conferences will reach full gender equality in their speaker, moderators and panels (similar number of male and female participants, about 50-50%)."* (From Government Policies, European Quantum Flagship, EU, 2020)

The "Responsible Technology Development" narrative calls for ethical development practices and highlights QT's ability to address critical global issues. It often references potentially harmful societal consequences such as misinformation, polarizing online speech, and discrimination caused by biased algorithms-concerns often extrapolated from past experiences with AI-that could be exacerbated by QT. Additionally, the narrative typically points to QT's role in achieving sustainable development goals, such as increasing renewable energy production. Consider the following representative examples:

*"To meet the increasing societal demands for trustworthy technology, the future of quantum technologies should be informed by lessons learned from previous digital ethics failures (for example, from instances where individuals' data was misappropriated or misused, where algorithms were employed to amplify misinformation and hateful content online, and where marginalized groups faced added discrimination through the use of unrepresentative data sets and unfair models)."* (From Business & Industry, Ernst & Young, 18.05.2022)

*"Quantum computing can optimize wind turbine and solar panel designs, increasing energy efficiency and further reducing our dependence on fossil fuels."* (From News Coverage, The Economic Times, 03.04.2023)

The "National Tech Strategies" theme focuses on the initiatives and policies adopted by governments to advance and establish leadership positions in QT and related fields such as semiconductor manufacturing. It features discussions on national research funding, government-sponsored quantum strategies and partnerships. In contrast to the "Politics & Global Conflict" theme, it spotlights the domestic aspirations of governments for QT development.

A commonplace narrative related to this theme is the "National Ambition and Leadership" narrative. Here, the debate focuses on plans and active management of QT capabilities to establish national leadership roles. The overriding priorities for building QT capabilities articulated in this narrative are primarily twofold: enhanced national security and economic interests. However, the narrative also links the pursuit of broader societal values, such as fairness and social inclusion, to QT. The following excerpts from U.S. and Australian policy documents illustrate as much:

*"Finally, it is imperative that while developing the QIS (quantum information science) enterprise in the United States, the Government also protects intellectual property and economic interests, seeks to understand dual-use capabilities, and supports national-security-relevant applications."* (From Government Policies, National Strategic Overview for Quantum Information Science, United States, 2018)

*"Vision for Australia's quantum future: In 2030, Australia is recognized as a leader of the global quantum industry, and quantum technologies are integral to a prosperous, fair and inclusive Australia."* (From Government Policies, National Quantum Strategy, Australia, 2023)

An additional prevailing narrative is that of "Collaboration and Cooperation", which emphasizes the importance of forming international and multi-sectoral partnerships. This narrative asserts the need for collaboration with industry, the security sector, and within multilateral fora such as the OECD and the G7. At its core, it advocates for the sharing of talent and knowledge across academia, industry, and government to foster an ecosystem for innovation. In contrast to the "National Leadership and Ambition" narrative, which focuses on establishing national leadership in QT, this narrative emphasizes the benefits of global collaboration and shared progress.

*"Continued and reinforced international collaborations as well as collaborations with industry and security sector."* (From Government Policies, Israel, 2018)

The "Business & Market Development" theme covers topics related to corporate strategies, private investment, and commercialization of QT. It also describes aspects of companies' recent technological breakthroughs and the challenges of bringing QT to market.

A key narrative is the "Investment and Market Potential" narrative. It presents QT as an investment opportunity, highlighting the sizable investments being made as well as the technology's potential to disrupt industries and create new markets. It often revolves around corporations such as Alibaba, Microsoft, and Google, along with startups such as D-Wave, Rigetti, and IonQ, which are attracting investment and driving the development and commercialization of QT:

*"Alibaba last year pledged to spend more than $15 billion in research and development through its Alibaba Damo Academy, which will bankroll frontier research like quantum computing."* (South China Morning Post, 23.04.2018)

*"Announced private investments for 2021 are already double this amount, bringing the total private investment in quantum computing from 2001 to 2021 to more than $3.3 billion."* (From Business & Industry, McKinsey & Company, 14.12.2021)

Relatedly, a regularly recurring storyline emphasizes the importance and urgency of actively engaging with QT today to realize its full potential tomorrow. We call this the "Getting Quantum Ready" narrative. It acknowledges the current immaturity and uncertainty surrounding QT but presents the technology as a promising frontier. This narrative specifically prompts business stakeholders to prepare for QT's eventual maturation, to avoid being caught off guard by the competitive pressures of QT innovation, and to capitalize on the opportunities it presents.

### 4.2. How do narratives compare across domain sources?

To quantify our observations, Figure 1 presents a heat map showing the percentage distribution of themes across the data sources. Figure 1 shows that Media coverage is evenly distributed across themes. The main focus is on "Technical Aspects & Applications" (28.3%), "National Technology Strategies" (23.8%), and "Politics & Global Conflicts" (19.5%). "People & Society" (15.9%) and "Business & Market Development" (12.5%) are also present, but to a lesser extent. Overall, the media covers a wide range of themes and narratives and reports on QT from multiple perspectives. Business reports primarily concentrate on "Technical Aspects & Applications" (66.6%), followed by "Business & Market Development" (22.2%). Other themes are scarcely covered. In stark contrast to the technical aspects, business reports barely touch on themes and narratives about "National Tech Strategies" (1.5%) or "Politics & Global Conflicts" (0.2%). The scope of businesses reports is thus limited to immediate business interests about leveraging the technology for commercialization potential rather than its broader political implications. Government texts give overwhelming attention to "National Tech Strategies" (39.6%) and "Technical Aspects & Applications" (37.8%). This indicates a strong intent on incorporating QT into policymaking, and a parallel interest in developing technical applications. This focus may be driven by an interest in regulation, but certainly also has its origins in the relevance of the technology for national security and defense purposes. Other themes, such as "Business & Market Development" (13.1%), "People & Society" (8.6%), and "Politics & Global Conflicts" (1.0%), receive less emphasis.

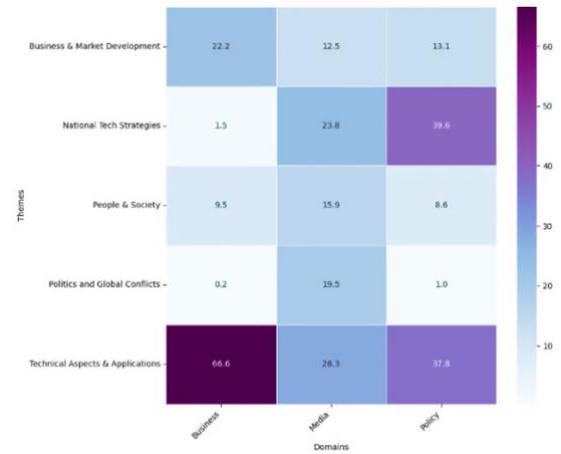

**Figure 1. Heat Map of Theme Distribution by Data Source (%).**

### 4.3. How has QT discourse changed over time?

Across the political, business, and media sectors, Figure 2 reveals noteworthy shifts in QT discourse over time. The most prominent trend is an increasing focus on the political implications of QT. This is reflected in the growing emphasis on "National Tech Strategies" in both the policy and media spheres, as well as the increasing attention to "Politics & Global Conflicts" in media coverage. This shift suggests a move away from purely technical discussions towards greater consideration of geopolitical and strategic implications. In the Policy sector, "National Tech Strategies" as well as "Technical Aspects & Applications" are the dominant themes across the period. Other themes in the policy sector, such as "People & Society" and "Politics & Global Conflicts", maintain relatively stable and low levels of attention throughout the timeframe. The business sector shows a different pattern. "Technical Aspects & Applications" is the primary focus, with pronounced peaks throughout. "Business & Market Development" follows a general upward trend, followed by a decline in recent years. Other topics, such as "National Tech Strategies" and "People & Society," maintain low but consistent levels of coverage throughout the period. Media coverage exhibits the most dynamic changes among the three sectors. There's a slight decrease in focus for "Technical Aspects & Applications" and a more pronounced decrease for "People & Society". As noted above, this shift is offset by an increase in attention to political issues. "Business & Market Development" maintains a relatively constant level of attention throughout the period.

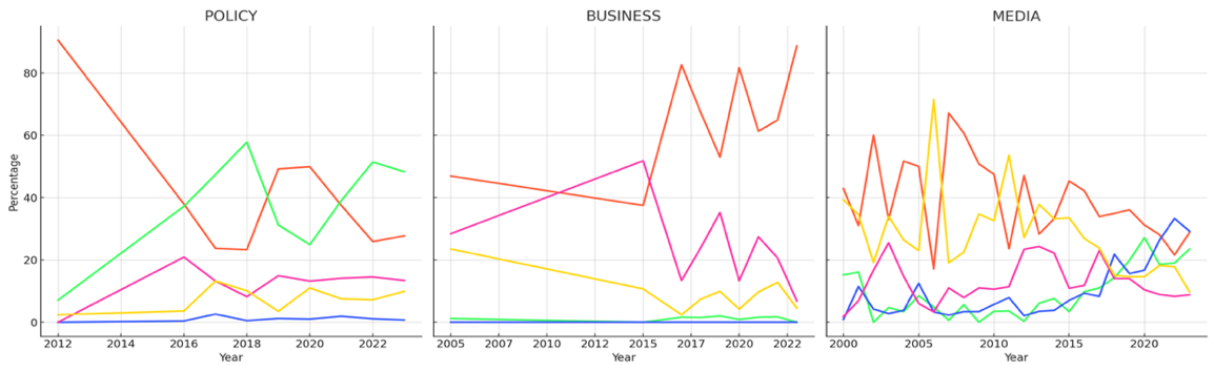

**Figure 2. Theme Distribution over Time (%).**

## 5. Discussion

Three key findings emerged from our analysis that deserve special attention. First, the overall public discourse centers on explaining and making sense of the basic concepts and principles of QT. Key narratives emphasize the potential of QT to leapfrog classical computing technologies and describe promising applications, similar to findings of previous studies (Godoy-Descazeaux et al., 2023; Meinsma et al., 2023). This focus is evident in the prominence of the "Technical Aspects & Applications" theme across the business, media, and political spheres, as well as its consistent presence over time. Much of the debate therefore focuses on communication efforts to explain theoretical concepts, practical benefits of QT, and the rationale for its development.

Although technical aspects and applications are prominent in all domains, the business, media, and political spheres differ. Our second main finding highlights these divergences. First, we saw an absence of political issues in business discourse, and second, a comparatively low attention paid to social issues in both corporate and governmental discourses. Regarding the former, companies that neglect the political dimensions of QT, whether domestic strategies or geopolitical concerns, expose themselves to risk. With low political awareness, companies and their stakeholders may miss chances to influence regulatory frameworks, be unprepared for new regulations, and be unaware of significant public investments that affect business opportunities. This can lead to a narrow focus that results in missed opportunities or overconfidence. A possible reason for this finding could be that companies may avoid sensitive topics in outreach efforts to protect other interests, e.g., commercial aspects or funding. In addition, it is important to keep in mind that the findings are based on public outreach, not internal conversations.

Regarding the latter, the relative lack of attention to societal issues in both business and government discourse can also be detrimental. QT is unlikely to be available directly to individual customers or citizens as a technical infrastructure. Instead, due to its technical complexity, it will likely be distributed via cloud solutions. If the public perceives important societal concerns, e.g., accessibility or diversity and inclusion, are not being taken seriously, this neglect can lead to low public confidence in QT and hinder its application to solve tasks for which it has great potential. In short, both companies and governments can benefit from paying attention to these blind spots when communicating to the public about QT.

Our third major insight concerns the parallels between AI and QT narratives. AI narratives often oscillate between exaggerated hopes and fears. This dichotomy has influenced national policies, media coverage, and public perceptions, resulting in a gap between AI's imagined potential and its actual capabilities. In contrast, our analysis shows that QT narratives do not fall into the utopian-dystopian spectrum often observed in AI discourse. While many narratives emphasize the potential changes QT could bring to industries and society, with media coverage progressively focusing on its political implications, extreme predictions of either catastrophe or salvation are decidedly absent. However, an increasing focus on narratives about politics and global conflict may become critical. Overemphasis on such narratives could fuel ongoing geopolitical competition, potentially hindering international collaboration and openness.

## 6. Conclusion & Outlook

Our analysis shows that the discourse surrounding QT is layered, with a range of narratives emerging in the business, media, and government sectors. The overall discussion is dominated by technical aspects and applications, which highlight QT's potential to transform industries and society. But there are noticeable gaps in the treatment of political and societal issues. Business narratives often overlook geopolitical and regulatory dimensions. Similarly, both business and government narratives tend to underemphasize the societal implications of QT. Comparing QT to AI narratives, QT discourse lacks the extreme speculative predictions, both pessimistic and optimistic, that often characterize AI-related discussions. Furthermore, in terms of future research, our approach is informed by quantitative methods and therefore provides a broad yet generalized overview of the QT discourse. To extend and complement these findings, future studies could employ qualitative methods to provide more in-depth and contextually informed insights into specific themes within the QT discourse.